\documentclass[ 10pt, conference]{IEEEtran}
\def\BibTeX{{\rm B\kern-.05em{\sc i\kern-.025em b}\kern-.08em
 T \kern-.1667em\lower.7ex\hbox{E}\kern-.125emX}}
\usepackage{graphicx}
\usepackage{amsmath}
\usepackage{mathrsfs}
\usepackage{mathtools, cuted}
\usepackage{amssymb}
\usepackage{multicol}
\usepackage{enumitem}
\usepackage{algorithm}
\usepackage{algpseudocode}
\usepackage{pifont}
\usepackage[english]{babel}
\usepackage{mathtools}
\usepackage{booktabs,caption,fixltx2e}
\usepackage[flushleft]{threeparttable}
\newcounter{MYtempeqncnt}
\begin{document}
\title { D2D User Selection For Simultaneous Spectrum Sharing And Energy Harvesting } 
\author{\begin{tabular}{cc}
 &\textbf{ Mansi Peer, Vivek Ashok Bohara, Neha Jain } \\
 & Wirocomm Research Group \\
 & Department of Electronics and Communication\\
 & IIIT-Delhi, India 110020\\
& Email: mansip@iiitd.ac.in, vivek.b@iiitd.ac.in, nehaj@iiitd.ac.in\\
 \end{tabular}
 }
\maketitle
\thispagestyle{empty}
\begin{abstract}
This paper presents a device-to-device (D2D) user selection protocol wherein multiple D2D pairs coexist with a cellular network. The D2D users help in extending the cellular services to a cellular user with worse channel conditions. In the developed framework, certain D2D users harvest energy and share the spectrum of the cellular users by adopting a hybrid time switching and power splitting protocol. The D2D user which achieves the desired target rate for the cellular communication  is selected to serve as a decode-and-forward (DF) relay for the cellular user. The proposed work analyzes the impact of increase in the number of D2D users on the performance of cellular user as well as derives an upper bound on the time duration of energy harvesting within which best possible rate for cellular user can be obtained. The performance of the proposed protocol has been quantified by obtaining the closed form expressions of outage probability. 
\end{abstract}
\section{Introduction}
It has been predicted that by the year 2020, 50 billion devices will be interconnected through a wireless network \cite{1001}. Such large scale  increase in the number of devices will lead to surge in spatial density of the devices. As a consequence, there would be plethora of devices which would like to communicate with each other. This motivated the researchers to propose a device-to-device (D2D) communication framework which is more  spectral and energy efficient than the conventional cellular communication. In a typical D2D communication, cellular users living in close proximity can form a direct link for data transmission without routing it through the core network or base station (BS) \cite{6805125}, \cite{5208020}. Since the link distances between the two D2D users are usually small, so are the energy requirements. Hence the possibility of powering the D2D devices through RF energy harvesting (also known as RF powered D2D devices) can not be ignored \cite{6951347}.
Energy harvesting (EH) from radio frequency (RF) signals has been proposed as a viable solution to alleviate the problem of non-renewable energy storage hence making the system ``green" and energy efficient. In \cite{6552840}, authors have proposed two energy harvesting protocols for a cooperative wireless network, namely the Time switching-based relaying (TSR), where  the energy constrained amplify and forward (AF) relay node switches between the energy harvesting and information decoding modes, and Power splitting-based relaying (PSR), where a fraction of power received at relay is devoted to energy harvesting and rest of the power is utilized for information decoding. The above two protocols were limited as fixed amount of time or fixed fraction of power received was devoted for energy harvesting and information decoding. 

Further, there are two broad spectrum access paradigms in D2D : (i) Inband D2D, where D2D and cellular users share the same spectrum band (ii) Outband D2D, where D2D users access the unlicensed band or the spectrum orthogonal to cellular users \cite{6805125}. A lot of recent work has been devoted to the area of inband D2D. The authors in \cite{5450264},\cite{6554553} proposed various methods of combating the issue of interference in inband D2D. In this paper we have assumed an inband D2D paradigm where the D2D users coexist with the cellular users, and interference is mitigated by using orthogonal time access in a two-phase hybrid protocol as discussed later. 

The proposed work utilizes two use-cases of D2D communications incorporated in Long term evaluation (LTE)-Advance standard : (i) Multihop relaying in cellular network, (ii) local voice and data services \cite{6231164}. We have assumed that the cellular communication gets hampered due to a link failure and the cellular network will undergo dual-hop relaying using the D2D devices. Further, the selected paired D2D users will carry out their own communication too.

We have employed a hybrid power splitting and time switching EH relaying protocol at the RF-powered D2D devices \cite{7536840}. This alleviates the drawbacks of previously proposed protocols \cite{6552840} as there is unequal division of time between the two phases of energy harvesting and transmission respectively. Further by utilizing this hybrid protocol it is possible to obtain an upper bound on the time duration of energy harvesting within which best possible quality of service (QoS) of cellular communication can be achieved. In addition to above, we also incorporate D2D user selection wherein the D2D user which harvests the maximum energy as well as achieves the desired target rate for the cellular user is selected to serve as a DF relay for the cellular user. As a consequence, only the orthogonal channel (channel corresponding to the selected D2D user) is required for cellular network's information transmission thus achieving the same outage performance as a conventional cooperative diversity technique using multiple orthogonal channels \cite{5397898}.

Some of the major contributions of the proposed work has been summarized as below:
\begin{itemize}
    \item A hybrid EH protocol for RF powered D2D devices is proposed that provides useful insights on the impact of unequal division of transmission slot on the QoS for cellular and D2D users.
    \item The D2D user and cellular user share the same spectrum and interference is mitigated by transmitting the data following orthogonal time access.
    \item The best D2D user selection ensures a certain QoS for the cellular and / or D2D user. The aim is to enhance cellular user's performance while providing a D2D pair an opportunity to share spectrum resources  allocated to the cellular user. To the best of our knowledge no work has been done in the area of EH based D2D user selection for spectrum sharing but the same has been addressed in the proposed algorithm.
\end{itemize}
The rest of the paper is organized as follows. Section II discusses the proposed system model and protocol, section III presents the outage analysis of the cellular and D2D users, section IV showcases the simulation results obtained and section V concludes the paper.

Throughout the paper, $E[.]$ denotes the expectation  and an exponentially distributed random variable $\psi$ with mean $\frac{1}{\lambda}$ is denoted by $\psi\sim\mathcal{E}(\lambda)$. A complex Gaussian random variable $\varrho$ with mean $\kappa$ and variance $\sigma^{2}$ is denoted by $\varrho\sim\mathcal{C}\mathcal{N}(\kappa,\sigma^{2})$.
\section{System Model and Protocol Description }
\begin{figure}
\centering
\includegraphics[width = 9cm, height = 75mm]{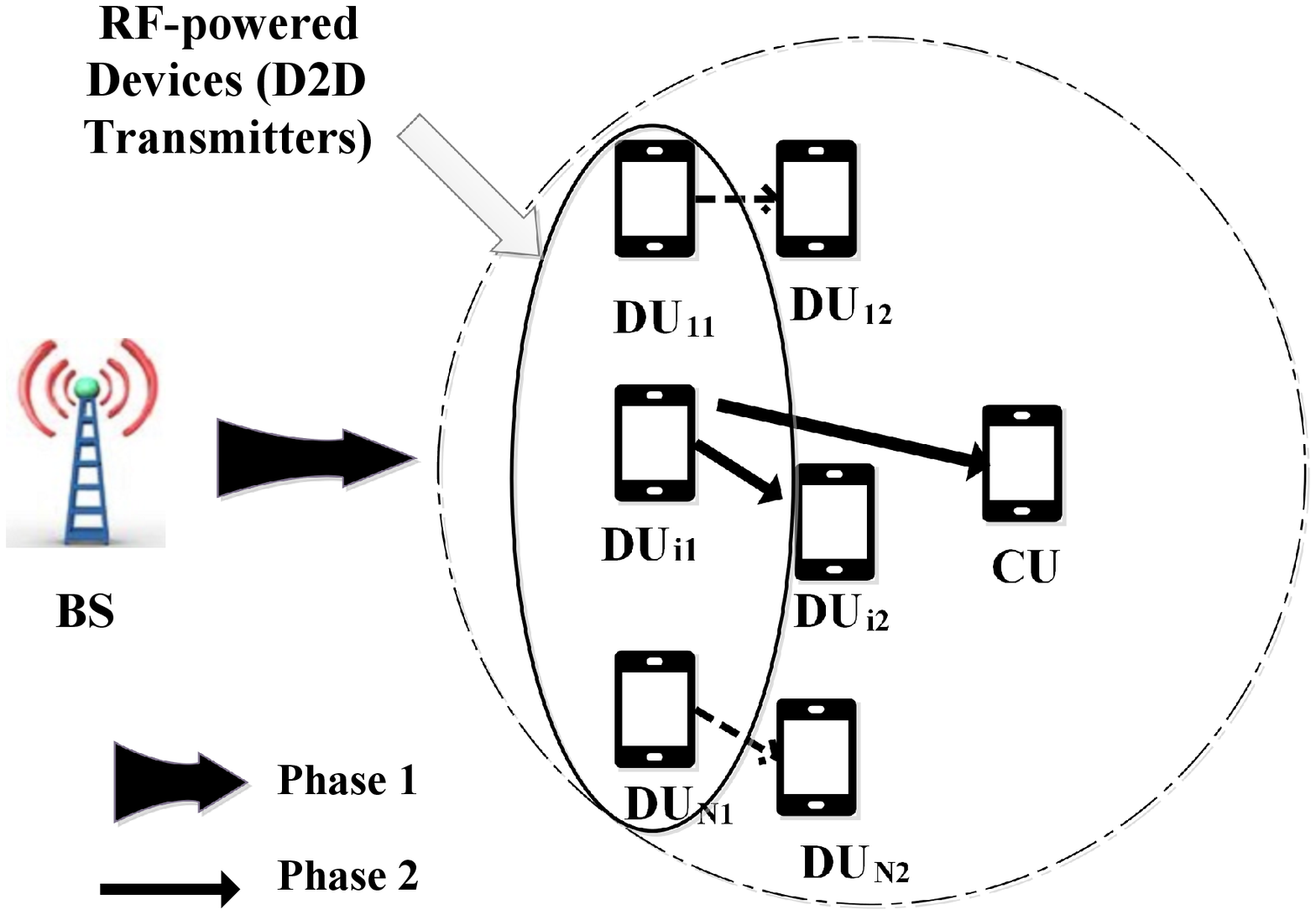}
\caption{System model where $DU_{i1}$ is the best D2D user.}
\end{figure}
 The proposed system model consists of base station $BS$, a cellular user $CU$, $N$ number of D2D pairs consisting of D2D transmitters ($DU_{i1}$) and receivers ($DU_{i2}$) as shown in Fig.1. It is assumed that each cellular user has been allocated a specific time frequency resource (resource blocks ) which is orthogonal to other cellular users. However, this resource can be shared among the cellular and D2D users. The channel between each pair of transmitter and receiver is assumed to be Rayleigh flat fading and is denoted by $h_{i1 }$ for $BS$-$DU_{i1}$ link, $h_{i2}$ for $DU_{i1}$-$CU$ link, $h_{i3}$ for $DU_{i1}$-$DU_{i2}$ link, $h_{i4}$ for $BS$-$DU_{i2}$ link, where $i$ = 1, 2, 3....$N$. Here, $h_{ij}\sim\mathcal{C}\mathcal{N}(0,d_{ij}^{-v})$  where $v$ is the path loss exponent, $d_{ij}$ is the distance between the respective transmitter and receiver and $j$ = 1, 2, 3, 4. The channel gain  $\beta_{ij}$ =  $|h_{ij}|^2$ and is denoted as $\beta_{ij}$ $ \sim\mathcal{E}(d_{ij}^{v})$. Further it is assumed that  $h_{ij}$  are independent and identically distributed (i.i.d.) $\forall$ $i$, $j$. Hence,  $d_{1}$ = $d_{i1}$, $d_{2}$ = $d_{i2}$, $d_{3}$ = $d_{i3}$ and $d_{4}$ = $d_{i4}$  $\forall i$. The additive white Gaussian noise (AWGN) at each receiver is denoted as $n$$\sim\mathcal{C}\mathcal{N}(0,\sigma^{2})$ with zero mean and $\sigma^{2}$ variance. 
 
 We consider that the transmitted signals from $BS$ to $CU$, $x_{c}$, and between the D2D pair, $x_{d}$, are zero mean with unit power i.e. $E[x_{c}]=E[x_{d}]= 0$ and $E[|x_{c}|^{2}]=E[|x_{c}|^{2}]= 1$. The signal received at $DU_{i1}$, $CU$, $DU_{i2}$ in $k^{th}$ phase  is denoted as $y^{k}_{du_{i1}}$, $y^{k}_{2}$ and $y^{k}_{du_{i2}}$ respectively.\\
 \begin{algorithm}[h]
\caption{Selection Procedure}
\begin{algorithmic}
\State Let $R_{du_{i1}}$  be the rate achievable at $DU_{i1}$ in phase 1, $R_{ct}$ is the target rate for cellular user and $R_{i}$ be the rate achievable at $CU$ when  $DU_{i1}$ is transmitting $x_{c}$ and $x_{d}$ with $\rho$$P_{h}$ and (1-$\rho$)$P_{h}$ power respectively, where $P_{h}$ is the harvested power at $DU_{i1}$. Also, $R_{du_{i2}}$ is the rate achievable at $DU_{i2}$ in phase 1.

\For{$i=1:1:N$}
 \If {$R_{du_{i1}} > R_{ct}$} 
\State Select $DU_{i1}$.
\EndIf
\State  $i\leftarrow i+1$
\EndFor
\State Create a decoding set $D$ of the selected D2D users.

\If {$D$ = $\emptyset$ }
 \State \textbf{case} { 1 }
\Else
  \If {$\max_{i}$ [$R_{i}] > R_{ct}$}
\State  Select the user from D for transmission of combination of $x_{c}$ and $x_{d}$ signal in phase 2.
          \If {$R_{du_{i2}} > R_{ct}$} 
           \State \textbf{case} { 2 }
              \Else
             \State \textbf{case} { 3 }
           \EndIf
   \Else
   \State \textbf{case} { 4 }
   \EndIf 
 \EndIf
\end{algorithmic}
\end{algorithm}
We assume that there is no direct link between $BS$ and $CU$ \cite{6552840}, hence $BS$ requires the assistance of the D2D user for transmission of the information $x_{c}$ to $CU$ and the selected D2D pair also gets an access to the spectrum. It is worth noting that BS constantly monitors the channel conditions and gathers all the necessary CSI for selecting the best D2D user. After this BS broadcasts the message which includes the information of selected D2D user and also the mode in which the selected D2D user has to operate. The D2D user selection process is carried out as given in Algorithm 1. The cases mentioned in the algorithm have been discussed later in the paper.

Energy harvesting is employed on the selected D2D transmitter and it follows a hybrid  power splitting and time switching protocol. The total transmission time $T$ is divided into two phases of unequal duration $\alpha$T and $(1-\alpha)T$ respectively, where $\alpha$ is the time switching factor. In phase 1 decoding of $x_{c}$ and energy harvesting occurs in parallel at the selected D2D transmitter. The power allocated for energy harvesting  and information decoding is $\gamma$$\mathcal{P}_{i}$ and (1-$\gamma$)$\mathcal{P}_{i}$ respectively, where $\mathcal{P}_{i}$ is the power received at $DU_{i1}$ and $0 \le \gamma \le 1$ is the power splitting factor\footnote{The hybrid protocol has been employed to a $i^{th}$ D2D transmitter to support our point that  prior analysis takes place at BS. However, during real time transmission only the selected D2D transmitter will follow hybrid protocol. }.

  In phase 1, $BS$ broadcasts its signal $x_{c}$  with a transmit power $P_{c}$. The signal received at  $DU_{i1}$ from  $BS$ is given by
 \begin{equation}
     y^{1}_{du_{i1}} = \sqrt{P_{c}}h_{i1}x_{c} +n.
 \end{equation}
  The signal received at the energy harvester branch of $DU_{i1}$ will be
\begin{equation}
    \sqrt{\gamma} y^{1}_{du_{i1}}=\sqrt{\gamma P_{c}}h_{i1}x_{c} +\sqrt{\gamma} n.
\end{equation}
Hence the energy harvested at each $DU_{i1}$ in time  $\alpha T$ can be given by 
\begin{equation}
E_{hi}= \eta\gamma{P_{c}}|h_{i1}|^2\alpha{T} 
\end{equation}
where $\eta$ $\epsilon$ (0,1] is the conversion efficiency of the RF to DC conversion circuitry used. The transmission power available at the energy harvesting D2D user is given by
\begin{equation}
    P_{hi} = \frac{E_{hi}}{(1-\alpha)T}=\frac{\eta\gamma{P_{c}}|h_{i1}|^2\alpha}{(1-\alpha)}.
\end{equation}
The signal received at  information receiver of each $DU_{i1}$ is given by
\begin{equation}
 \sqrt{(1-\gamma )}y^{1}_{du_{i1}}=\sqrt{(1-\gamma)P_{c}}h_{i1}x_{c} + n.
 \end{equation}

Hence the rate achievable at $DU_{i1}$ will be

  \begin{equation}
   R_{du_{i1}}=\alpha \log_{2}(1+ SNR_{du_{i1}})
   \end{equation}
   where $ SNR_{du_{i1}}$ = $\frac{(1-\gamma)P_{c}|h_{i1}|^2}{\sigma^{2}}$.
   
   Similarly, signal received at $DU_{i2}$ in phase 1 will be
  \begin{equation}
  y^{1}_{du_{i2}}=\sqrt{P_{c}}h_{i4}x_{c} + n.
  \end{equation}
  
 Hence the rate achievable at $DU_{i2}$ in phase 1 will be
 \begin{equation}
  R_{du_{i2}}= \alpha \log_{2}\left(1+\frac{P_{c}|h_{i4}|^2}{\sigma^{2}}\right).
 \end{equation}
 
 For phase 2 transmission, four cases (or modes) are possible and they are illustrated in Table I. The four cases signify the possible achievable rates for $DU_{b1}$-$DU_{b2}$ link depending on the channel conditions and amount of energy harvested.

  \section{Outage Analysis}
 \subsection{Cellular Communication System}
The probability that no $DU_{i1}$ is able to decode $x_c$ in phase 1 is given by
\begin{equation}
P_{1} = P[R_{du_{11}} < R_{ct}]P[R_{du_{21}} < R_{ct}]......P[R_{du_{N1}} < R_{ct}].
\end{equation}
It is assumed that each  $DU_{i1}$ is equidistant from  $BS$ and $CU$. Hence,
\begin{align} \nonumber
P_{1} = \left(P[R_{du_{11}} < R_{ct}]\right)^{N},\\
= \left(1- e^{\frac{-d^{v}_{1}t}{m(1-\gamma)}}\right)^{N},
\end{align}

where $t = (2^\frac{R_{ct}}{\alpha} -1)$ and $m = \frac{P_{c}}{\sigma^{2}}$. 

The probability that at least one of the D2D transmitters decode $x_c$ will be
\begin{equation}
P^{'}_{1} = 1 - P_{1}.
\end{equation}

Thus the outage for the cellular system can be given as (assuming $k$ D2D transmitters have successfully decoded $x_c$)

\begin{equation}
P_{oC} = P_{1} + P^{'}_{1}P_{2}, 
\end{equation}
 where 
 \begin{equation}
     P_{2} =\left[\sum_{k=1}^{N}\binom{N}{k}p^{N-k}(1-p)^{k}P[\max(R_{1}...R_{k}) < R_{ct}]\right], 
 \end{equation}
 \begin{equation}
     R_{i} = (1-\alpha)\log_{2}\left(1+\frac{\rho P_{h}|h_{i2}|^2}{(1-\rho)P_{h}|h_{i2}|^2 +\sigma ^{2}}\right),
 \end{equation} 
  
  \begin{equation} 
  P[\max(R_{1}...R_{k}) < R_{ct}] =
\begin{cases}\left[1-uK_{1}(u)\right]^{k},&\text {$\alpha<1-\delta$}\\
1,&\text{ otherwise}
\end{cases},
\end{equation}
  
  where $u = \sqrt{\frac{4a}{b d^{-v}_{1}d^{-v}_{2}}}$,
  
$a=\frac{\sigma^{2}(1-\alpha)(2^{R_{ct}/(1-\alpha)}-1)}{\eta\gamma P_{c}\alpha} $,

$b=\rho-(2^{R_{ct}/(1-\alpha)}-1)(1-\rho)$,

$\delta =\frac{R_{ct}}{\log_2(1+\frac{\rho}{1-\rho})}$ and $K_{1}(.)$ is the modified first order bessel function of second kind \cite{6552840}. The closed form expression for cellular outage obtained on substituting (13), (14), (15) in (12) will be given by
 \begin{equation}
 P_{oC} = 
 \begin{cases}
 P_{1}+ P^{'}_{1}\left[\sum_{k=1}^{N}\binom{N}{k}p^{N-k}(1-p)^{k}w^{k}\right] ,&\text{$ \alpha < 1-\delta$}\\
P_{1}+ P^{'}_{1}\left[\sum_{k=1}^{N}\binom{N}{k}p^{N-k}(1-p)^{k}\right] = 1,&\text{ otherwise}
 \end{cases},
\end{equation}
where $w = \left[1-uK_{1}(u)\right]$ and $p$ = $\left(1- e^{\frac{-d^{v}_{1}t}{m(1-\gamma)}}\right)$.

\begin{table*}
 \begin{tabular} { |c|p{4.1cm}|l|l|l| }
  \hline
  Case & Condition & Signal received at $DU_{b2}$ &  Rate achievable corresponding to $DU_{b1}$-$DU_{b2}$ link \\ [5.5pt]
  \hline
   Case 1 & a) Decoding set $D$ is empty. \newline b) D2D transmitter achieving maximum  rate at its paired D2D receiver corresponding to $x_{d}$ transmission is selected.  & $y^{2}_{du_{b2}} =  \sqrt{P_{h}}h_{b3}x_{d} +n$ & $R_{b3}=(1-\alpha)\log_{2}\left(1+\frac{P_{h}|h_{b3}|^2}{\sigma^{2}}\right)$  \\ [12pt]
  \hline
        
  Case 2 & a) Decoding set $D$ is non empty. \newline b) A D2D user in $D$ meets the criterion $max_{i}$$[R_{i}]$ \textgreater $R_{ct}$. \newline c) $DU_{b2}$ decodes $x_{c}$ successfully in phase 1. & $y^{2'}_{du_{b2}}=\sqrt{(1-\rho)P_{h}}h_{b3}x_{d} + \sqrt{\rho P_{h}}h_{b3}x_{c} +n$ &  $R^{'}_{b3}=(1-\alpha)\log_{2}\left(1+\frac{(1-\rho)P_{h}|h_{b3}|^2}{ \sigma^{2 }}\right)$\\ [12pt]
  \hline            
  Case 3 & a) Decoding set $D$ is non empty. \newline b) A D2D user in $D$  meets the criterion $max_{i}$$[R_{i}]$ \textgreater $R_{ct}$. \newline c) $DU_{b2}$  unsuccessful in decoding $x_{c}$ in phase 1.  & $y^{2''}_{du_{b2}}=\sqrt{(1-\rho)P_{h}}h_{b3}x_{d} + \sqrt{\rho P_{h}}h_{b3}x_{c} +n$  & $R^{''}_{b3} = (1-\alpha)\log_{2}\left(1+\frac{(1-\rho)P_{h}|h_{b3}|^2}{ \rho P_{h}|h_{b3}|^2 + \sigma^{2 }}\right)$ \\ [12pt]
  \hline 
  Case 4 & a) Decoding set $D$ is non empty. \newline b) No D2D user in $D$ meets the criterion $max_{i}$$[R_{i}]$ \textgreater $R_{ct}$.
   \newline c) A D2D user in $D$ achieving maximum rate at its paired $DU_{i2}$ corresponding to $x_{d}$ transmission  is selected. & $y^{2'''}_{du_{b2}} =  \sqrt{P_{h}}h_{b3}x_{d} +n$ & $R^{'''}_{b3}=(1-\alpha)\log_{2}\left(1+\frac{P_{h}|h_{b3}|^2}{\sigma^{2}}\right)$  \\ [12pt]
   \hline
  \end{tabular}
  \begin{tablenotes}[H]
      \small
      \item Note: $DU_{b1}$ and $DU_{b2}$ constitutes the selected D2D pair. $h_{b3}$ denotes the channel coefficient of the link between $DU_{b1}$ and $DU_{b2}$. $P_{h}$ is the power harvested at $DU_{b1}$.
    \end{tablenotes}
    \caption{Table containing the details of the four possible cases in phase 2.}
\end{table*}
\subsection{Device-to-Device Communication System}
The outage of the D2D system can be obtained by combining the four cases mentioned in Table I,
 \begin{multline}
P_{oD} = P_{1}P[\max(R_{13}....R_{N3})<R_{dt}] + \\
 P^{'}_{1}[1-P_{2}]P[R^{'}_{b3}<R_{dt}]P[R_{du_{b2}}>R_{ct}] +  \\
P^{'}_{1}[1-P_{2}]P[R^{''}_{b3} < R_{dt}]P[R_{du_{b2}} < R_{ct}] + \\
 P^{'}_{1}P_{2}\left[\sum_{k=1}^{N}\binom{N}{k}p^{N-k}(1-p)^{k}P[\max(R_{13}...R_{k3})< R_{dt}]\right],
 \end{multline}

where $R_{dt}$ is the D2D target rate and $R_{i3} $ is the rate corresponding to $DU_{i1}$-$DU_{i2}$ link when only D2D signal is transmitted by $DU_{i1}$.
Further,
\begin{equation}
    P[R^{'}_{b3} < R_{dt}] = \left[1-vK_{1}(v)\right],
\end{equation}
where v = $ \sqrt{\frac{4c}{d^{-v}_{1}d^{-v}_{3}}}$ and c = $ \frac{\sigma^{2}(1-\alpha)(2^{R_{dt}/(1-\alpha)}-1)}{\eta\gamma P_{c}\alpha(1-\rho)}.$

\begin{equation}
    P\left[ \max(R_{13}R_{23}....R_{i3}) < R_{dt}\right] = {\left[1-yK_{1}(y)\right]}^i.
\end{equation}
\begin{equation}
      P[R^{''}_{b3} < R_{dt}] =
    \begin{cases}
    1-zK_{1}(z),&\text{ $\alpha < 1- \mu$}\\
    1,&\text{otherwise}
    \end{cases},
    \end{equation}
where $\mu=\frac{R_{dt}}{log_{2}\left(\frac{1}{\rho}\right)}$,
$y=\sqrt{\frac{4d}{d^{-v}_{1}d^{-v}_{3}}}$
$z=\sqrt{\frac{4d}{ed^{-v}_{1}d^{-v}_{3}}}$,

$d=\frac{\sigma^{2}(1-\alpha)(2^{R_{dt}/(1-\alpha)}-1)}{\eta\gamma P_{c}\alpha}$ and $e=(1-\rho)- \rho\left(2^\frac{R_{dt}}{1-\alpha}- 1 \right)$. 
\begin{equation}
    P[R_{du_{i2}} > R_{ct}] = e^{\frac{-d^{v}_{4}t}{m}}.
    \end{equation}
     By substituting (10), (15), (18), (19), (20) and (21) in (17) we get (22) where
$Q = \left[1-yK_{1}(y)\right]$, $R = \left[1-vK_{1}(v)\right] $,\\ $T = \left[\sum_{k=1}^{N}\binom{N}{k}p^{N-k}(1-p)^{k}w^{k}\right]$, \\$P_{21} = P_{1}+ P^{'}_{1}\left[\sum_{k=1}^{N}\binom{N}{k}p^{N-k}(1-p)^{k}w^{k}\right]$, \\$P_{22} = P_{1}+ P^{'}_{1}\left[\sum_{k=1}^{N}\binom{N}{k}p^{N-k}(1-p)^{k}\right] $ and $\phi = e^{\frac{-d^{v}_{4}t}{m}}$.
\begin{figure*}[!t]
\normalsize
\setcounter{MYtempeqncnt}{\value{equation}}
\setcounter{equation}{21}
\begin{align}
 P_{oD} =
 \begin{cases}
P_{1}Q^{N} + P^{'}_{1}(1-P_{21})R\phi + P^{'}_{1}(1-P_{21})T(1-\phi) + P^{'}_{1}P_{21}\left[\sum_{k=1}^{N}\binom{N}{k}p^{N-k}(1-p)^{k}Q^{k}\right] ,& \text{ $\alpha  <1-\delta$ and  $\alpha < 1- \mu$} \\
  P_{1}Q^{N} + P^{'}_{1}(1-P_{21})R\phi + P^{'}_{1}(1-P_{21})(1-\phi) + P^{'}_{1}P_{21}\left[\sum_{k=1}^{N}\binom{N}{k}p^{N-k}(1-p)^{k}Q^{k}\right],&\text {$\alpha < 1-\delta $ and  $\alpha \geq 1-\mu$} \\
 P_{1}Q^{N} + P^{'}_{1}(1-P_{22})R\phi + P^{'}_{1}(1-P_{22})T(1-\phi) + P^{'}_{1}P_{22}\left[\sum_{k=1}^{N}\binom{N}{k}p^{N-k}(1-p)^{k}Q^{k}\right] ,&\text {$\alpha \geq 1-\delta $ and  $\alpha < 1-\mu$}\\
 P_{1}Q^{N} + P^{'}_{1}(1-P_{22})R\phi + P^{'}_{1}(1-P_{22})(1-\phi) + P^{'}_{1}P_{22}\left[\sum_{k=1}^{N}\binom{N}{k}p^{N-k}(1-p)^{k}Q^{k}\right],&\text {$\alpha \geq 1-\delta $ and  $\alpha \geq 1-\mu$} 
\end{cases}.
\end{align}
\setcounter{equation}{\value{MYtempeqncnt}}
\end{figure*}
\begin{figure}
\centering
\includegraphics[width= 90mm, height = 65mm]{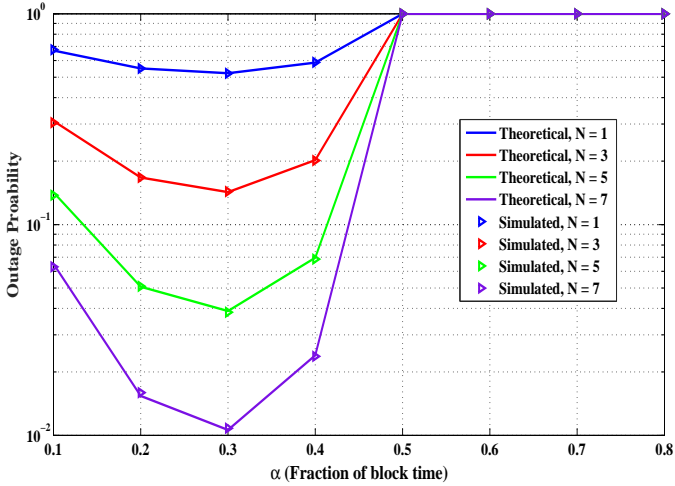}
\caption{Outage probability of the cellular network.}
\end{figure}
\begin{figure}
\centering
\includegraphics[width= 90mm, height = 65mm]{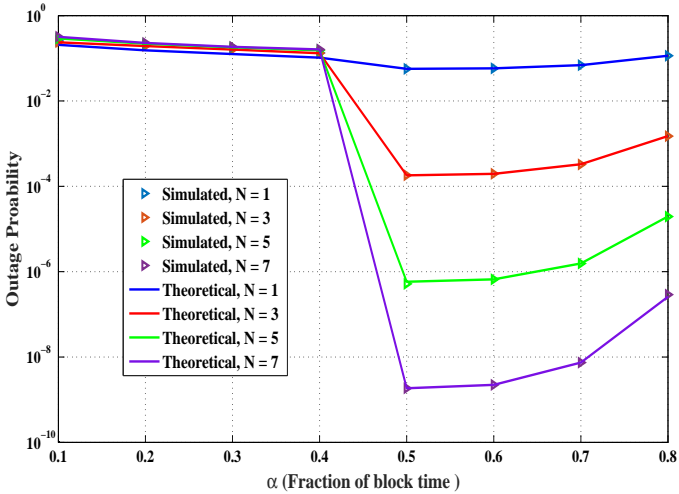}
\caption{ Outage probability of D2D network.}
\end{figure}

\section{Simulation Results }
This section simulates the effect of $\alpha$, $\rho$ and $N$ on the outage performance of the cellular and D2D network. Further in order to verify the analytical derivations, simulation results have also been compared with theoretical results. The distances between BS-$DU_{i1}$, $DU_{i1}$-$DU_{i2}$, $BS$-$DU_{i2}$ and $DU_{i1}$-$CU$ are assumed to be 30 m, 20 m, 10 m and 20 m respectively whereas distance between the $BS$-$CU$ link is 50 m. The remaining simulation parameters are given in Table II. 

\begin{table}[h]
 \caption{ Simulation Parameters}
 \centering
 \begin{tabular} { |c|c|c|c|c|c| }
  \hline
  Parameter & $\sigma^{2} $ & $\eta$ & $\gamma$ & $P_c$ & $v$\\ [7pt]
  \hline
 Value & -90 dBm & 0.8 & 0.75 & 10 dBm & 3 \\ [7pt]
  \hline
\end{tabular}
\end{table}

Fig.2 shows the variation in the outage probability for the cellular network with respect to $\alpha$ for $\rho$ = 0.75. It is quite obvious that the performance improves with N. This is due to the fact that decoding probability of $x_{c}$ in phase 1 increases with N (9). Further, corresponding to each value of $\rho$ there is a value of $\alpha$ beyond which increase in $N$ has no impact on the outage performance. This upper bound on the value of $\alpha$ for each value of $\rho$ has also been theoretically verified. For instance when $\rho$ = 0.75 it can be verified from (16) that  the upper bound is at $\alpha$ = 0.5. Beyond $\alpha = 0.5$ there is no direct link between $BS$ and $CU$ so when  $DU_{i1}$-$CU$ link fails $\forall$ $DU_{i1}$ $\epsilon$ D, cellular outage increases to 1. 

Fig.3 depicts the variation in the outage probability for D2D communication with respect to $\alpha$ for $\rho$ = 0.75. Fig.3 shows the trend similar as Fig.2 with respect to N.  This is due to the fact the proposed scheme selects the D2D transmitter which has the maximum harvested energy. Hence with an increase in number of D2D users there is more probability that the selected D2D user will have sufficient harvested energy to relay the information to the cellular user, $x_{c}$,  with less outage.  

\section{Conclusion}
In this paper a D2D user selection protocol for simultaneous energy harvesting and spectrum sharing in an underlying cellular system was presented. Specifically, a method of first select and then harvest has been adopted for best D2D user selection. Through the simulation and analytical results it was shown that by increasing the number of D2D users the performance of the cellular system can be significantly improved. Further the impact of power allocation factor $\rho$ and time switching factor $\alpha$ on the outage performance was also investigated. Closed form analytical expressions for the cellular and D2D outage probabilities were derived and validated using the simulation results.

\nocite{*}
\bibliographystyle{IEEEtran}
\bibliography{Selection_D2D.bbl}
\end{document}